\documentclass[onecolumn]{article}
\usepackage{authblk}
\usepackage[utf8]{inputenc}
\usepackage{siunitx}
\usepackage{graphicx}
\usepackage{subcaption}
\usepackage{placeins}
\usepackage[usenames,dvipsnames,svgnames,table]{xcolor}
\usepackage[linkcolor=black]{hyperref} 
\usepackage[colorinlistoftodos,textsize=tiny]{todonotes}
\title{Restoring betatron phase coherence in a beam-loaded laser-wakefield accelerator}
\author[1]{A.~Koehler}
\author[1]{R.~Pausch}
\author[1,2]{M.~Bussmann}
\author[1]{J.P.~Couperus~Cabada{\u{g}}}
\author[1]{A.~Debus}
\author[1]{J.M.~Krämer}
\author[1,3]{S.~Sch{\"o}bel}
\author[1]{O.~Zarini}
\author[1,3]{U.~Schram}
\author[1]{A.~Irman}
\affil[1]{Helmholtz-Zentrum Dresden -- Rossendorf, Bautzner~Landstrasse~400, 01328~Dresden, Germany}
\affil[2]{Center for Advanced Systems Understanding, 02826~G\"orlitz, Germany}
\affil[3]{Technische Universit\"at Dresden, 01062~Dresden, Germany}

\date{\today}

\begin{document}

\maketitle

\begin{abstract}
Matched beam loading in laser wakefield acceleration (LWFA), characterizing the state of 
flattening the accelerating electric field along the bunch, leads to the minimization of energy spread at
high bunch charges. 
Here, we experimentally demonstrate by independently controlling injected charge and accelerating
gradients, using the self-truncated ionization injection scheme, that minimal energy spread
coincides with a reduction of the normalized beam divergence. 
With the simultaneous confirmation
of the micrometer-small beam radius at the plasma exit, deduced from betatron radiation spectroscopy, we attribute this effect to the minimization of chromatic betatron decoherence.
These findings are supported by rigorous three-dimensional particle-in-cell simulations tracking self-consistently particle trajectories from injection, acceleration until beam extraction to vacuum. 
We conclude that beam-loaded LWFA enables highest longitudinal and transverse phase space densities.
\end{abstract}

\section{Introduction}
The concept of laser wakefield acceleration (LWFA) 
exploits ultra-high accelerating field gradients
of up to a few hundred gigavolt-per-meter generated 
in the wake of a high-intensity laser pulse as it propagates through an optically transparent plasma~\cite{Esarey2009,Downer2018}.
Electron bunches can thus be accelerated to GeV energies within centimeters~\cite{Gonsalves2019}.
Beam quality with respect to bunch charge, energy bandwidth,
emittance and pulse-to-pulse stability has improved
substantially  during the last decade and is closely linked to a variety of
controlled electron injection techniques~\cite{Geddes2008,Osterhoff2008,Rechatin2009,Gonsalves2011,Buck2013,Irman2018}.
Only recently it was demonstrated that laser-plasma accelerators
can be tailored for minimum energy spread at high bunch charges
by reshaping the local accelerating field via matched beam loading~\cite{Irman2018,Couperus2017,Goetzfried2020,Kirchen2021}.
This combination of high charge, essential for the beam loading
regime, and the short bunch duration in the range of \SI{10}{\femto\second}~\cite{Glinec2007,Debus2010,Lundh2011,Zarini2018,Zarini2021}
results in high peak-current beams exceeding \SI{10}{kA}.
Future applications,
such as high-field THz sources~\cite{Green2016},
laboratory-size beam-driven plasma accelerators~\cite{Hidding2010,MartinezDeLaOssa2013,Kurz2021}
and compact free-electron lasers (FELs)~\cite{Schlenvoigt2008,Fuchs2009,Maier2012,Huang2012,Steiniger2014},
will benefit from such compact and further improved high-brightness electron sources.

\begin{figure}[h]
 \centering
    \includegraphics[width=0.6\textwidth]{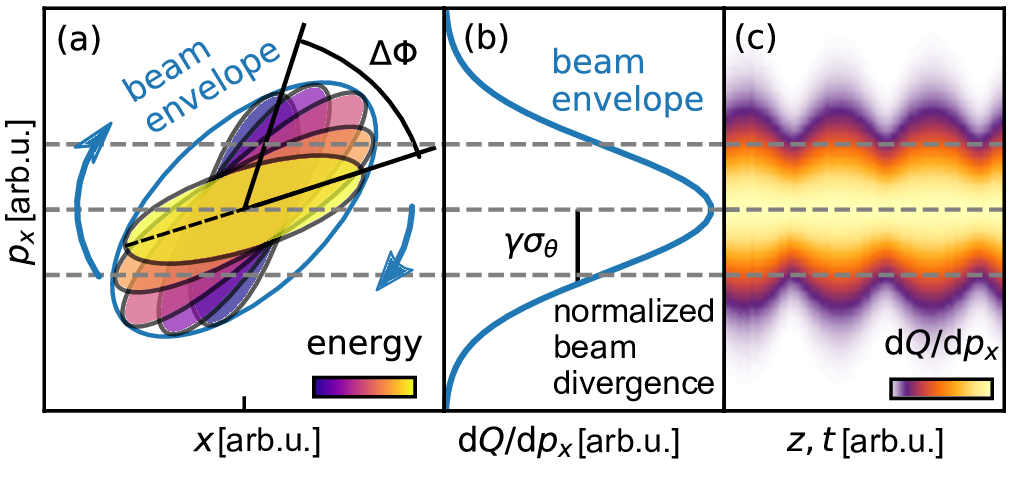}
    \caption{Beam decoherence: 
	    (a) illustrates the transverse phase space of a bunch with finite energy spread 
	    consisting of many slices of different electron energy which rotate in phase space 
        with the energy-dependent betatron frequency
        directed by the focusing force.
	    The beam envelope encircles the sum of all slices.
	    (b) shows the projection $\mathrm{d}Q/\mathrm{d}p_x$ of the beam on $p_x$,
	    which is typically recorded in experiments and used for determining the normalized beam divergence $\gamma\sigma_\Theta$.
	    (c) shows the time-dependence that is caused by the phase-space rotation of the beam envelope.
	    When the slices in (a) span over $\Delta \phi\ge\pi$, full decoherence is reached
	    and the modulation in (c) vanishes.
	    For $\Delta \phi <\pi$, the beam can be extracted at a phase with reduced momentum spread and small $\sigma_\Theta$.
	    }
    \label{fig:illustration-decoherence}
\end{figure}
For realizing such high-quality beams,
not only beam energy spread 
but also transverse emittance, dominated by beam divergence~\cite{Eichner2007,Andre2018}, has to be minimized.
In transverse phase space  $(x,p_x)$, 
a polychromatic bunch can be divided into slices representing different energy classes, schematically illustrated in figure~\ref{fig:illustration-decoherence}. 
The wakefield provides a strong linear focusing force~\cite{Khachatryan2007}, such that
off-axis electrons perform transverse (betatron) oscillations around the beam axis and emit radiation
while being accelerated~\cite{Esarey2002,Albert2008,Schnell2012,Corde2013a,Downer2018}. 
Due to the coupling of longitudinal acceleration and linear focusing, the relativistic mass increase of the electrons results in energy-dependent angular frequencies, amplitudes, and phases. 
Hence, the beam undergoes significant envelope oscillations (figure~\ref{fig:illustration-decoherence}(c))~\cite{Mehrling2012,Xu2014}.
As a consequence, the betatron oscillations can eventually loose phase coherence for a polychromatic beam.
For energy-chirped bunches typical in LWFA,
a phase difference $\Delta\phi\ge\pi$ between the highest and the lowest energy slice leads to full decoherence
and maximizes the occupied phase-space area, i.e., in beam divergence and size.
Only for $\Delta\phi<\pi$, the phase space is not filled entirely and thus the beam can be coupled out either at minimum beam size or divergence.
Simulation studies suggest that a minimum phase difference is only achievable for a very short injection duration that results in both a small initial phase spread and a similar accelerating field thus leading to a small final energy spread~\cite{Mehrling2012,Xu2014}. 
This, however, has the disadvantage of only providing  small charges of a few picocoulomb.

Here, we report on experimental findings in the beam loading regime that show improved transverse beam quality by restoring beam coherence while providing a high charge.
Despite having hundreds of picocoulomb charges, control on the beam coherence decreases beam divergence by about \SI{20}{\percent}.
At the matched beam loading condition, 
a balance between the bunch's self-fields and the accelerating field of the wakefield is reached, 
resulting in a constant longitudinal accelerating field along the bunch~\cite{Katsouleas1987,Tzoufras2008}. 
As a result, all electrons within the bunch experience the same accelerating field 
so that no energy bandwidth is added during the acceleration process. 
Completing the examination of $(x,p_x)$,
betatron radiation spectroscopy confirms a beam radius of less than \SI{0.7}{\micro\meter} at the plasma exit.
The simultaneous measurements of the small beam size and the minimized divergence demonstrate 
an increased transverse beam quality and reduce emittance, facilitating beam transport and novel applications.
Since beam loading can reduce the energy spread,
the control of this spread provides a unique opportunity to study electron beam coherence.
For further investigating the coherence-restoring process close to the experimental conditions,
we performed systematic high resolution, self-consistent particle-in-cell (PIC) simulations that confirmed our findings.
To completely study the coherence restoration, we examined electron ionization, trapping and acceleration from low to high energies in detail.  
The effect of space charge during field-free propagation to the detector system is assessed by tracing particles with a particle tracker code.
The combination of experimental and simulation data provides valuable insight into the evolution of the phase space and thus
enhances the understanding of coherence and decoherence in high-charge LWFA.

\section{Experimental setup}
\begin{figure}[h!]
    \centering
      \includegraphics[width=0.6\textwidth]{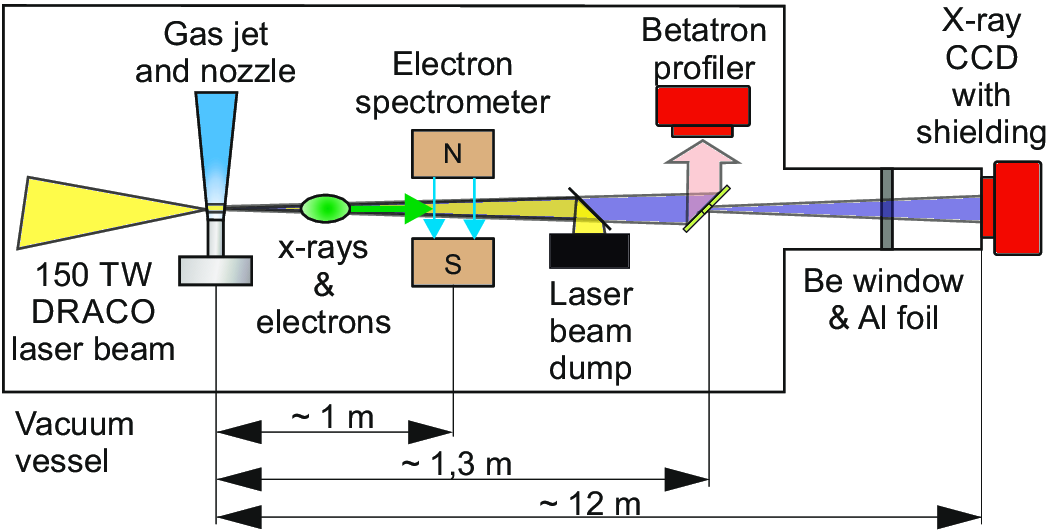}
    \caption{Experimental setup:
      The laser is focused on a gas jet and drives a wakefield.
      Accelerated electrons are energy-analyzed in a  magnet spectrometer.
      An aluminum foil reflects residual laser light to a beam dump behind the spectrometer.
      The scintillator of the betatron profiler intercepts the angular profile of the betatron radiation.
      On-axis betatron radiation passes through a hole in the profiler and is detected by the x-ray camera, 
      which is separated from the interaction chamber by a beryllium window.
      An aluminum filter foil attenuates the betatron flux.}
    \label{fig:experimentalsetup}
\end{figure}
Experiments were performed with the DRACO laser system~\cite{Schramm2017} at the HZDR.
Figure~\ref{fig:experimentalsetup} shows a schematic of the setup reproducing conditions as in~\cite{Couperus2017}.
Laser pulses of \SI{30}{f s} full-width at half-maximum (FWHM) duration with \SI{2.5}{J} energy on target
were focused with an off-axis parabolic mirror (f/20) to a vacuum focal spot size of \SI{20}{\micro m} (FWHM)
yielding a normalized laser intensity of $a_0\simeq 2.6$.
The Strehl ratio of the wavefront-corrected and transversely symmetric beam was measured to be \num{0.9}.
The laser beam was focused \SI{\sim 1.5}{\milli\meter} beyond the center of a \SI{3}{\milli\meter} long de Laval gas nozzle.
The nozzle  was 
operated with He-N$_2$ gas mixtures containing \SIrange{0.2}{1.5}{\percent} of N$_2$.
The gas density profile at \SI{1.5}{\milli\meter} above the nozzle exit,
i.e., the laser beam axis,
exhibits a \SI{1.6}{\milli\meter} flat-top region~\cite{Couperus2015}.
Electrons were injected into the wakefield using self-truncated ionization injection (STII)~\cite{Couperus2017,Zeng2014,Mirzaie2015},
which enabled stable and reproducible shots with high charges of up to \SI{500}{\pico\coulomb} (within FWHM).
In order to measure the energy distribution and divergence, 
a \SI{40}{cm} long permanent magnet dipole, positioned downstream of the plasma accelerator, dispersed the accelerated electrons 
to a set of charge calibrated scintillator screens (Konica Minolta OG 400)~\cite{Kurz2018,Schramm2017}.

A back-illuminated, deep depletion x-ray CCD (Princeton Instruments Pixis-XO 400BR)
with \SI{1340 x 400}{pixel} recorded betatron radiation emitted from the LWFA process. 
The camera was placed inside a dedicated radiation-shielded area \SI{12}{m} downstream of the plasma target,
covering a solid angle of \SI{1.5x0.8}{\milli\radian}.
A \SI{76}{\micro m} thick beryllium window sealed the CCD chip, allowing for cooling and background noise reduction.
An aluminum filter foil of \SI{200}{\micro m} thickness attenuated the high betatron flux of up to \SI{20000}{photons \per eV} at the peak.
This attenuation was necessary to enable single-shot reconstruction of the betatron spectrum by counting single-photon events~\cite{Jochmann2013,Kohler2016,Kramer2018}.
Behind the electron spectrometer, 
a scintillator screen was located,
oriented at \SI{45}{\degree} with respect to the beam axis,
to record the angularly resolved betatron profile.
A \SI{5}{\milli\meter} aperture in this screen allowed the transmission of the on-axis betatron radiation towards the x-ray CCD camera.

\section{Experimental results}
\begin{figure}[h!]
 \centering
    \includegraphics[width=0.6\textwidth]{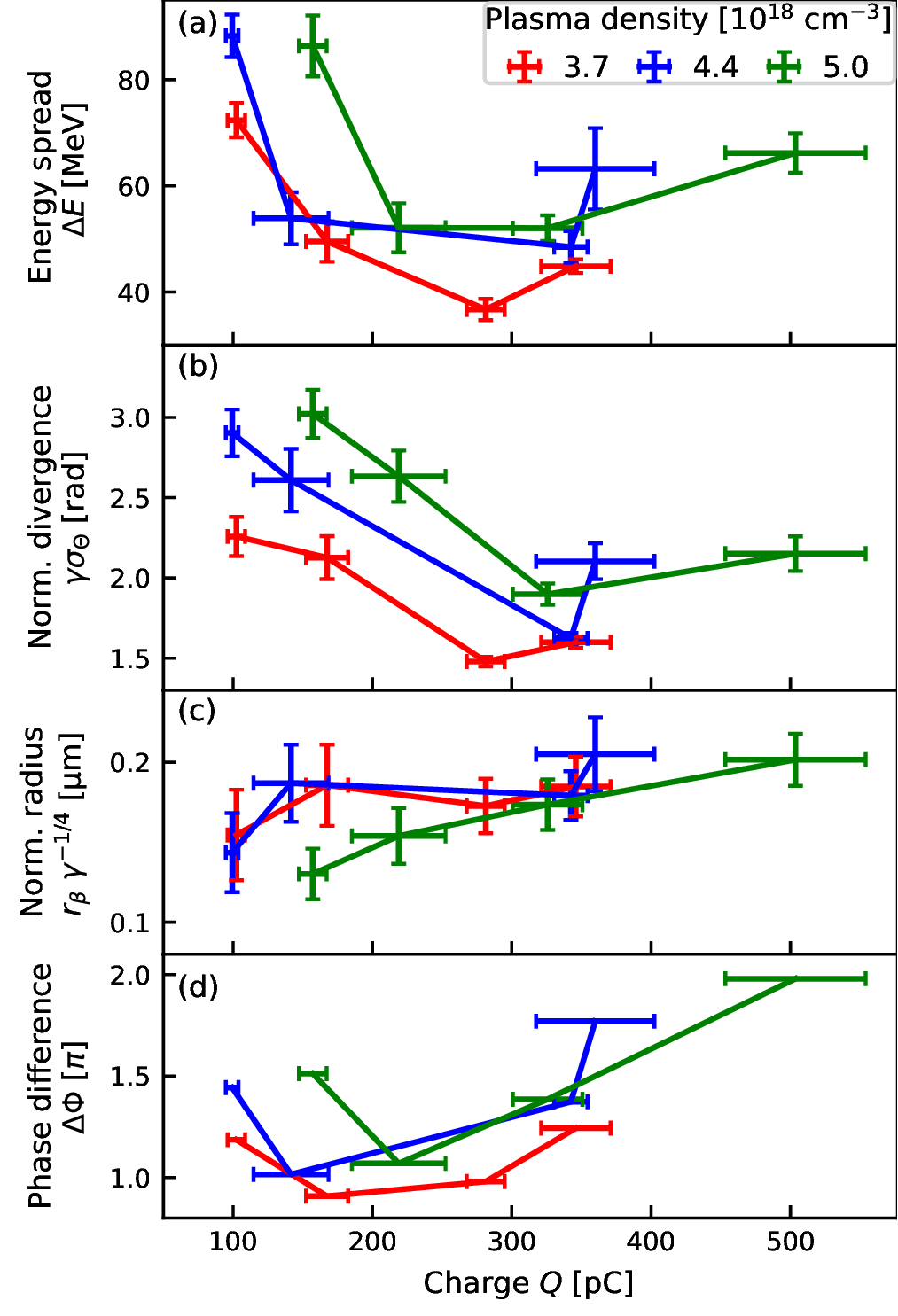}
    \caption{Measured energy spread $\Delta E$ (a), normalized divergence $\gamma \sigma_\Theta$ (b),
	    betatron radius $r_\beta\gamma^{-1/4}$ (c) and 
	    calculated betatron phase difference $\Delta\phi$ according to equation~(\ref{eq:phase-difference}) (d) obtained from electron bunches with different charges
	    and at three different plasma densities, 
	    \SI{3.7e18}{\per\cubic\centi\meter} (red line), \SI{4.4e18}{\per\cubic\centi\meter} (blue line) and \SI{5.0e18}{\per\cubic\centi\meter} (green line).
	    Every data point represents the average of up to $15$ shots with constant experimental parameters and
	    the error bars denote the standard error of the mean.
	    The error bars in (c) include the systematic error obtained from sensitivity analysis~\cite{Debus2010}.
	    The maximum beam energy ranges from \SIrange{300}{450}{MeV}.
	    }
    \label{fig:experimental-results}
\end{figure}

To access the beam loading regime in a controlled manner,
the amount of injected charge was tuned by varying the nitrogen concentration,
while keeping laser parameters and plasma density constant~\cite{Irman2018}.
Applying this approach, the location, volume and duration of the injection do not alter significantly between various dopings.
Thus the electron dynamics during injection, acceleration and extraction are predominantly influenced by the beam charge and the consequent local modification of the accelerating field, as also supported by simulations.
In the data analysis for electron and betatron spectra,
only shots were evaluated that showed a narrow-band energy feature and were oriented on-axis. 
Hence, we excluded spectrally modulated electron beams~\cite{Glinec2008} and off-center shots
that may originate from asymmetric plasma waves~\cite{Popp2010,Mittelberger2019} induced by higher-order transverse fluctuations of the laser beam or shot-to-shot pointing jitter.
Electron beam parameters such as maximum energy and bandwidth, divergence, and beam size at the accelerator exit are studied as a function of injected charges for a set of three plasma densities.

The effect of beam loading on the suppression of the accelerating field~\cite{Tzoufras2008} is evident in energy $E$ and energy bandwidth $\Delta E$ of the accelerated electron bunches.
While $E$ decreases from \SIrange{400}{300}{MeV} with increasing charge (see appendix~\ref{app:experimental-figures}), 
$\Delta E$ reduces reaching
a minimum at \SI[multi-part-units = single]{300\pm 50}{\pico\coulomb}, as presented in figure~\ref{fig:experimental-results}(a).
This charge value is referred to as matched beam loading~\cite{Couperus2017} for our experimental parameters. 
Injection of less or more charges from this optimum results in field gradients
that can lead to a broadening of the bunch energy bandwidth.

In order to investigate the transverse electron dynamics,
the geometric divergence of the beam $\sigma_\Theta$ is extracted  
from the non-dispersive plane of the electron spectrometer after having left the plasma accelerator.
Defocusing effects of the magnetic fields are corrected during data analysis.
Figure~\ref{fig:experimental-results}(b) shows the normalized beam divergence $\gamma\sigma_\Theta$ 
for various plasma densities
where $\gamma \approx E/(m_e c^2)$ is the relativistic Lorentz factor, $m_e$ and $c$ are the electron mass and the speed of light, respectively.
In particular, a similar charge-dependent trend as in $\Delta E$ is clearly seen.
At matched beam loading,
a drop of up to \SI{20}{\percent} of the normalized beam divergence is measured alongside the minimal energy bandwidth. 
This consistently occurs
for all plasma densities, which suggests that the accelerator geometry, 
i.e., cavity size, plasma wavelength, and the plasma down-ramp length, 
does not play a dominant role in the measured divergence reduction. 
In addition, deduced from the measured betatron spectrum~\cite{Schnell2012,Kohler2016}, the betatron source size and thus the electron beam radius  
is compared for the measured charges at the end of the accelerator.
Figure~\ref{fig:experimental-results}(c) presents
the reconstructed betatron source radii $r_\beta \gamma^{-1/4}$, where
the $\gamma^{-1/4}$ factor normalizes differences due to the relativistic acceleration~\cite{Downer2018}.
Note that the betatron radius exhibits no extremum at \SI{300}{\pico\coulomb},
but maintains a micrometer-small source size.
For this spatially compact bunch,
the normalized divergence reduces at the matched beam loading charge.
This indicates that space charge does not dominate the beam divergence
during acceleration in the plasma
as well as during propagation in the drift space, as confirmed below by simulations.
Otherwise, a monotonic increase of the beam divergence with charge should be expected.

In order to explain the reduced divergence observed at matched beam loading, we investigate the beam decoherence.
The degree of decoherence is defined by the maximum difference of the betatron phase $\Delta \phi = \phi_\textrm{max} -\phi_\textrm{min}$.
As shown in figure~\ref{fig:illustration-decoherence}, an electron in a slice $i$ with energy $\gamma_i$ rotates in the transverse phase space $(x,p_x)$ with the energy-dependent betatron frequency $\omega_{\beta,i}=\omega_p /\sqrt{2\gamma_i}$ where 
$\omega_p$ is the plasma frequency.
It thereby gains the phase $\phi_i=\int ^{l_\textrm{acc}} _0 \text{d}t\,\omega_{\beta,i}$ over the acceleration length $l_{\textrm{acc},i}$.
Assuming this electron experiences a constant accelerating field $E_{z,i}$, it gains the energy $\gamma_i = \gamma_0 + \gamma^\prime_{i} l_{\textrm{acc},i}$ with $\gamma^\prime_i = eE_{z,i} / m_e c^2$ denoting the Lorentz factor gain per distance.
The integration of $\phi_i$ then yields
\begin{equation}
    \phi_i = \frac{\sqrt{2}k_p}{\gamma^\prime_{i}} \sqrt{\gamma_0 + \gamma^\prime_{i}  l_{\textrm{acc},i}} + \phi_0,
    \label{eq:phase}
\end{equation}
where 
$k_p=\omega_p/c$, $\gamma_0$ and $\phi_0$ are the plasma wave number, initial electron energy and initial phase, respectively.
For full decoherence ($\Delta \phi\ge\pi$),
the injected electrons fill the entire phase space available during betatron oscillation and temporal minima in divergence can not be reached anymore. 
Assuming an injection duration much smaller than the inverse betatron frequency, 
reaching full decoherence
requires an acceleration length of $l_\textrm{dc}= 2.35 \lambda_\beta {\langle E\rangle}/{\Delta E}$~\cite{Mehrling2012,Michel2006},  with $\lambda_\beta = \sqrt{2 \gamma}\lambda_p$ being the betatron wavelength, 
$\lambda_p=2\pi c/\omega_p$ the plasma wavelength, 
and $\langle E\rangle$ the average energy of the peaked spectrum at the end of acceleration.
For the electrons observed in experiment, $l_\textrm{dc}$ is longer than two millimeters and thus longer than the acceleration length $l_\textrm{acc} = \SI{1.6}{\milli\meter}$.
Therefore, the accelerated electron bunches will not fully experience beam decoherence assuming an instantaneous injection.
When further assuming that electrons are injected at rest ($\gamma_0=1, \Delta \phi_0=0$) 
and that the energy spread is small  ($\Delta E \ll \left< E \right>$), an estimation of the phase difference  $\Delta \phi$ can be determined solely on values measured in the experiment~\cite{Xu2014}:
\begin{equation}
  \Delta \phi
    \approx \frac{\sqrt{2} k_p}{\gamma^\prime_{\textrm{max}}}\sqrt{1 + \gamma^\prime_{\textrm{max}} l_\textrm{acc} } 
          - \frac{\sqrt{2} k_p}{\gamma^\prime_{\textrm{min}}}\sqrt{1 + \gamma^\prime_{\textrm{min}} l_\textrm{acc} }.
  \label{eq:phase-difference}
\end{equation}
Here, $\gamma^\prime_{\mathrm{min/max}} \propto E_{z,\mathrm{min/max}}$, which are deduced from the measured energy gain and spread.
Representing equation~(\ref{eq:phase-difference}), figure~\ref{fig:experimental-results}(d) presents the phase difference $\Delta\phi$
as a function of charge, exhibiting a minimum in the vicinity of the matched beam loading condition as described by the charge corresponding to minimal energy spread.
This similarity in the behaviour of $\Delta\phi$ and beam divergence $\gamma\sigma_\Theta$ suggests that a reduction of the phase difference below the decoherence threshold ($\Delta\phi =\pi$) under optimal beam loading conditions could be responsible for the observed divergence minimum.
For a phase advance $\Delta\phi_0\ne 0$ accumulated during injection, highly relativistic energies (${\gamma}^\prime l_\textrm{acc}\gg 1$), and small relative energy spreads ($\Delta\gamma^\prime \ll \bar{\gamma}^\prime$), equation~(\ref{eq:phase-difference}) can be simplified to:
\begin{equation}
     \Delta\phi \approx \Delta\phi_0 - \Delta\phi_\text{beam}=
     \Delta\phi_0
     -\frac{k_p \Delta\gamma^\prime}{\sqrt{2}\bar{\gamma}^{\prime 2}} \sqrt{\bar{\gamma}^\prime l_\textrm{acc}}
     \label{eq:phase-difference-with-injection}
\end{equation}
with $\Delta\gamma^\prime=\gamma^\prime_\text{max}-\gamma^\prime_\text{min}$ and  $\bar{\gamma}^\prime=(\gamma^\prime_\text{max}+\gamma^\prime_\text{min})/2$.
Equation~(\ref{eq:phase-difference-with-injection}) indicates that $\Delta\phi_0$ can be compensated by careful tuning of energy and energy spread. 
It directly expresses the scaling of two main quantities of beam loading:
the $\bar{\gamma}^\prime$ is proportional to the mean accelerating field and $\Delta\gamma^\prime$ is proportional to the accelerating field gradient, which leads to energy spread during acceleration.
The scaling of equation~(\ref{eq:phase-difference-with-injection}) shows two trends. 
First, at injected charges far below beam loading, the mean accelerating field is at maximum. 
When increasing the injected charge towards beam loading, $\Delta\gamma^\prime$ and $\bar{\gamma}^\prime$ decrease. 
Second, while $\phi_\text{beam}$ vanishes for ideal beam loading ($\Delta\gamma^\prime=0$), the counteracting scalings of $\bar{\gamma}^\prime$ and $\Delta\gamma^\prime$ indicate the existence of a maximum in $\Delta\phi_\text{beam}$ before optimal beam-loading is reached.
The detailed phase progression with injected charge depends on the beam loading dynamics and
can be evaluated from experimental data on the final electron energy and energy spread varying with injected charge.

In the following section, the hypothesis of minimized phase spread will be further investigated.
In particular, the previous model assumptions will be expanded by dedicated simulations focusing on the injection dynamics and subsequent phase advance during acceleration.

\color{black}

\section{Simulations}
\label{sec:simulations}
For testing  the decoherence hypothesis,
we performed rigorous three-dimensional particle-in-cell (PIC) simulations with the PIConGPU code~\cite{Bussmann2013,PIConGPU041} 
modeling realistic experimental parameters (see appendix~\ref{app:PIC}).
This setup can be found under~\cite{PIConGPUSetup2020}.
As in the experiment, to inject more charge into the wakefield, the nitrogen doping concentration was increased for fixed laser-plasma parameters.
Moreover, a particle identification scheme implemented in the code allows for tracking the nitrogen K-shell electrons, source of the STII injected bunch, through the simulation. 
For beam propagation in vacuum towards a virtual 2D spectrometer, the General Particle Tracer (GPT) code~\cite{GPT} was deployed which is fed by the PIC particle data at the accelerator exit using an openPMD-based particle reduction \cite{openPMD,Ksenia2021}. 
This start-to-end simulation enables us to investigate the complete six-dimensional phase space dynamics in detail,
starting from electron injection and trapping, acceleration, until extraction and vacuum propagation.

\subsection{Injection}
\begin{figure}[h!]
    \centering
    \includegraphics[width=0.43\textwidth,height=6cm]{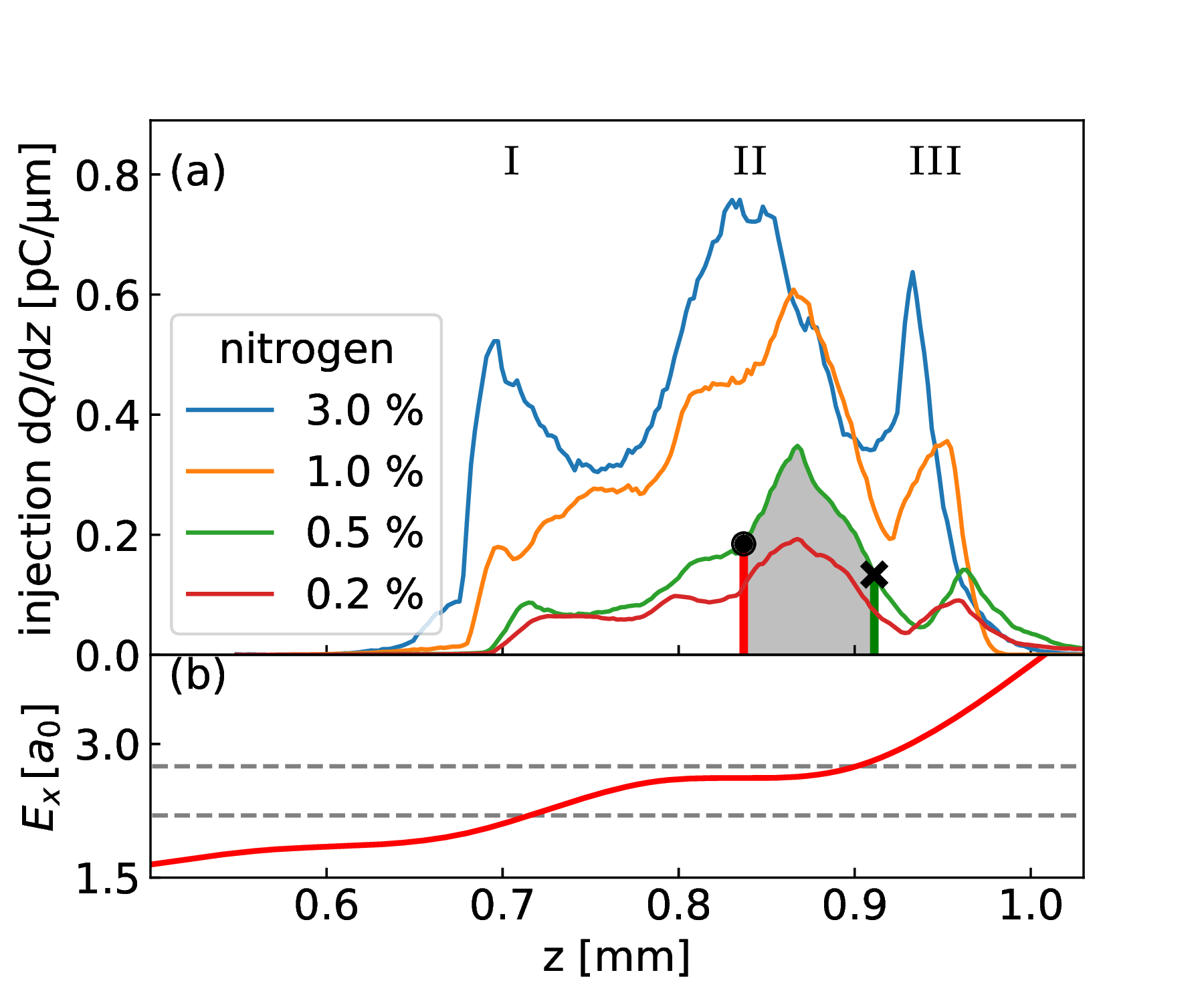}
    \includegraphics[width=0.43\textwidth,height=5.3cm]{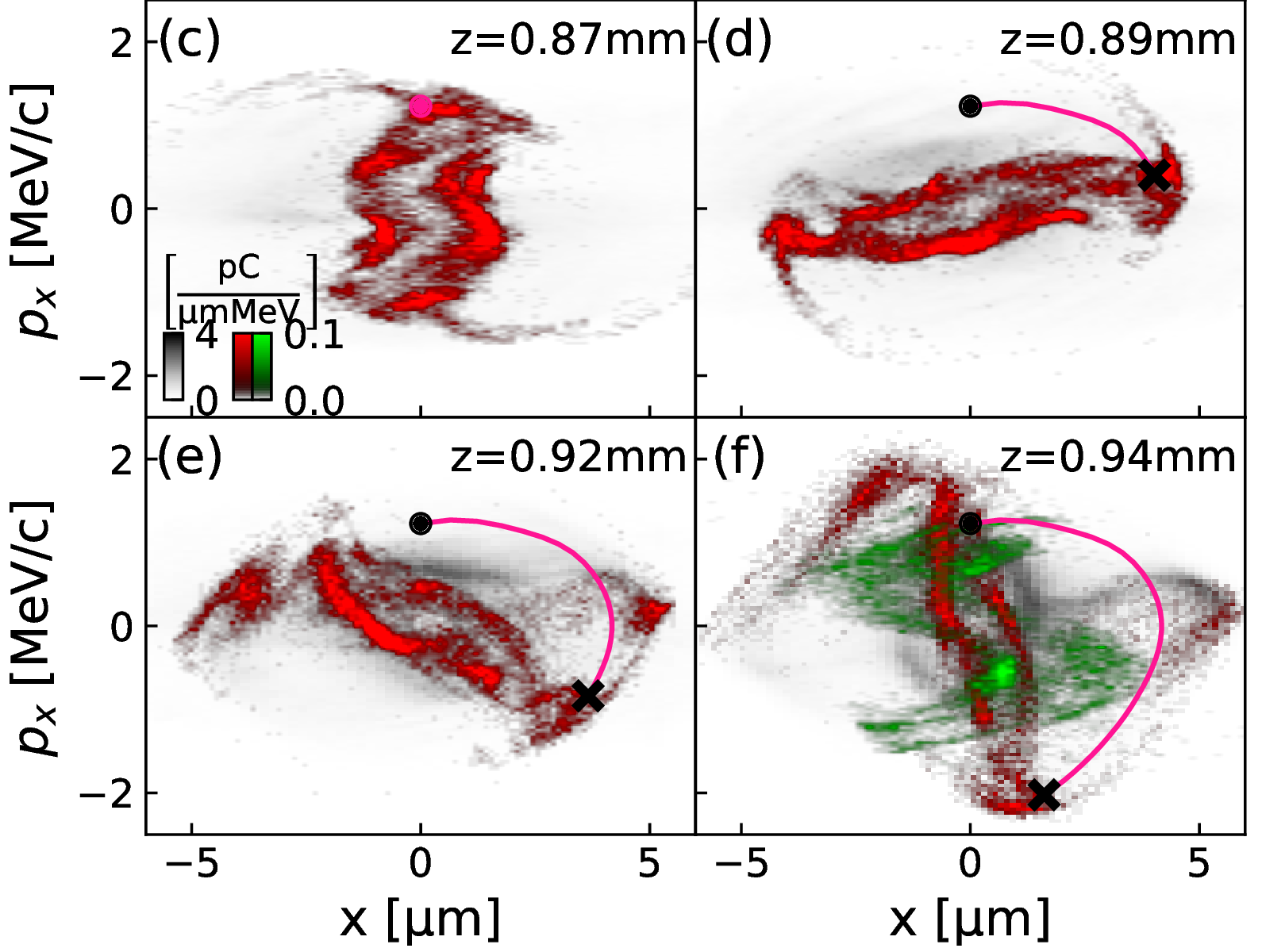}
    \caption{K-shell ionization and electron injection: 
                (a) shows the injection rate $\mathrm{d}Q/\mathrm{d}z$ of all trapped electrons and three distinct injections (I, II and III).
                (b) indicates the evolution of the laser peak intensity $a_0$.
                The gray dashed lines indicate the ionization threshold for the first and second nitrogen K-shell electrons.
                (c) to (f) presents the electron transverse phase space distribution $(x,p_x)$ in the laser polarization plane for the $0.5\%$ nitrogen doping at $z=$\SIlist{0.87;0.89;0.92;0.94}{\milli\meter}. 
                The gray scale represents macro-particles from region II marked in (a).
                The red and green scale represent slices of earliest
                and latest 
                injected electrons as indicated in (a), respectively.
                Extracted from the simulation, the red line represents the path of sample macro-particles
                that are injected (dot) in the earliest slice at (c) and rotation until the actual time (cross) shown in (d), (e) and (f).
                The simulations were performed for a plasma density of \SI{4.4e18}{\per\cubic\centi\meter}.
                }
    \label{fig:injection}
\end{figure}
Figure~\ref{fig:injection}(a) presents the injection rate, i.e. number of electrons trapped from the nitrogen K-shell as the laser propagates along the $z$-axis,
for nitrogen doping concentrations of $0.25\%$ (blue line), $0.5\%$ (orange line), $1.0\%$ (green line) and $3.0\%$ (red line). 
Once trapped in the wakefield, these electrons are accelerated to form a high energy and small bandwidth bunch at the accelerator exit, 
where the total accumulated charges of \SIlist{30;50;100;150}{\pico\coulomb} within FWHM energy bandwidth are generated from the lowest to the highest doping, respectively. 
At the end of the plasma channel, the macro-particles in the FWHM of the energy are selected and traced back to the injection to deduce $\textrm{d}Q/\textrm{d}z$.
While the total injection length is about \SI{300}{\micro\meter} for all dopings, three distinct regions of high injection rate $\textrm{d}Q/\textrm{d}z$ with an injection length of less than \SI{100}{\micro\meter} can be identified. 
Particularly, the majority of electrons is injected within the second region peaked at $z\approx\SI{0.84}{\milli\meter}$ and dominates the transverse beam dynamics.  
The  analysis of the betatron phase will focus on this group of electrons.

The $\textrm{d}Q/\textrm{d}z$ peaks are closely linked to the evolution of the laser peak intensity, as presented in figure~\ref{fig:injection}(b).
During propagation inside the plasma, the peak intensity of the laser pulse increases due to self-focusing which consequently impacts the shape of the wakefield. 
When the intensity exceeds the ionization threshold of the first nitrogen K-shell and the trapping condition is satisfied~\cite{Pak2010,Chen2012},  electrons from this shell are injected with the rate presented by the region I and II.
While the laser pulse further self-focuses,
the peak intensity increases and this eventually leads to the injection of the second K-shell electrons in the region III.
Injection self-terminates once the trapping condition is lost due to the wakefield evolution. 

Concentrating on region II, 
figure~\ref{fig:injection}(c) to (f) show the transverse phase space $(x,p_x)$ on the laser polarization plane for the $0.5\%$ doping during injection, 
i.e. at $z \approx\SI{0.87}{\milli\meter}$ to \SI{0.94}{\milli\meter}.  
Comparing the phase space of earliest injected electrons (red) to the last (green), 
the red slice has rotated by almost $\pi$ before the green slice was injected at the end of region II.
The injection duration leads to an initial phase-advance of $\Delta\phi\approx\pi$ and
thus acceleration starts with a fully decoherent bunch.
As an example, the magenta line indicates the path of a small sample of macro-particles traced by the simulation.
The same injection evolution also applies for other doping concentrations.

\subsection{Acceleration}
\begin{figure}[h!]
 \centering
     \includegraphics[width=0.85\textwidth]{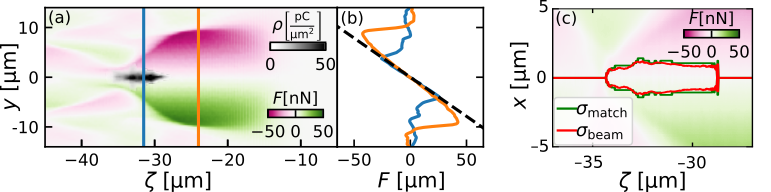}
        \caption{Beam loading in PIC simulation:
    (a) illustrates the focusing field map $F = e(E-cB)$ for an injected charge of \SI{100}{p C} at $z=\SI{1.7}{\mm}$.
    The density of injected electrons 
    and the focusing force are indicated by the gray and color scale at the bottom right, 
    respectively.
    (b) presents two line outs of the focusing force in front of the bunch (orange)
    and at the bunch (blue) emphasizing its linearity.
    The black dashed line in (b) indicates the theoretically predicted value $-E_0/2 k_p r$~\cite{Lu2007}.
    (c)  shows the beam size $\sigma_\textrm{beam}(\zeta)$  (red line) and 
    the matched beam size $\sigma_\textrm{match}(\zeta)$ (green line)
    after injection ($z=\SI{1.1}{\mm}$). 
    }
    \label{fig:focusing-fields}
\end{figure}
During acceleration, the focusing forces of the wakefield ensure that the beam stays close to the axis. 
Figure~\ref{fig:focusing-fields}(a) illustrates the focusing field map of the wakefield at $z=\SI{1.7}{\milli\meter}$
which is loaded with a relatively high charge of \SI{100}{\pico\coulomb}, generated by the $1\%$ nitrogen doping.
The blue and orange line outs in figure~\ref{fig:focusing-fields}(b) are sampling the focusing force $F$ at different longitudinal positions $\zeta=z-ct$,
i.e. in the front of the bunch and at the bunch center, respectively,
emphasizing the linearity and independence of $F$ on space charge.
Despite that the accelerating field being locally modified due to the beam loading, the focusing remains radially linear and longitudinally constant.
This important feature can preserve the transverse slice emittance for the acceleration of high-quality beams
and thereby allows to study the beam evolution, i.e., beam matching and decoherence.
In the case of beam matching, $F$ balances beam defocusing driven by finite emittance and
the beam size remains constant along the acceleration distance~\cite{Esarey2002}. 
An unmatched beam oscillates in beam size similar to a not fully decoherent bunch.
Figure~\ref{fig:focusing-fields}(c) shows the RMS beam envelope of the \SI{100}{pC} charge case confined within the wakefield focusing field $F$ at $\mathrm{z}=\SI{1.0}{\milli\meter}$. 
The red line represents the beam size $\sigma_{\mathrm{beam}}(\zeta)$ and the green line shows the matched beam size $\sigma_{\mathrm{match}}(\zeta)$ given by~\cite{Esarey2002} 
\begin{equation}
    \sigma_\textrm{match}(\zeta) = \gamma \left(\frac{ m_e c^2 }{\epsilon_n(\zeta) F(\zeta)} \right)^{1/3},
    \label{eq:beam-matching}
\end{equation}
where $\epsilon_n(\zeta)$ is the RMS normalized transverse emittance of the particle distribution at $\zeta$.
Even for large dopings as shown for the \SI{100}{pC} case in figure~\ref{fig:focusing-fields}(c), 
the beam matching condition is already satisfied at the beginning of the acceleration
and maintained during the adiabatic acceleration ($\sigma_\textrm{beam} \propto \gamma^{-1/4}$).
Together, the initially matched beam and the decoherent phase space distribution immediately after injection
allow us to study the beam decoherence during the subsequent acceleration process 
since the reversal of beam decoherence can be attributed solely to the acceleration gradient set by the beam loading. 

\begin{figure}[h!]
 \centering 
%
%
    \includegraphics[width=0.75\textwidth]{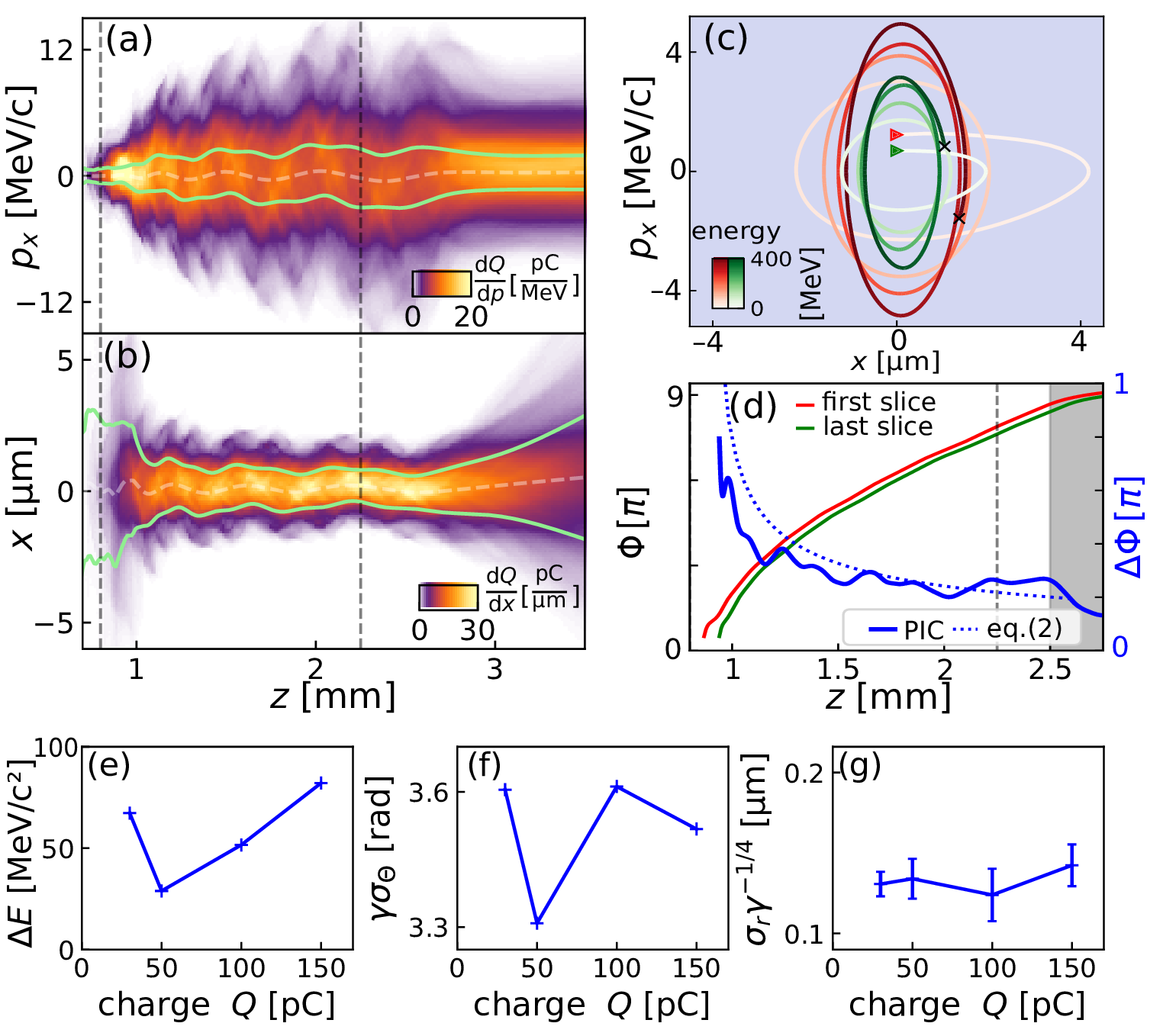}
    \caption{PIC particle tracking for \SI{50}{\pico\coulomb} charge,
	    (a) and (b) showcase the evolution of transverse momentum $p_x$ and coordinate $x$, respectively.
	    The green line indicates one standard deviation $\sigma_{p} (z)$ and $\sigma_x (z)$.
	    The vertical gray dashed lines indicate the plasma density up- and down-ramp.
        In (c), the red and green line represent particle orbits in $(x,p_x)$ from the earlier and later injected slice in figure~\ref{fig:injection}(c).
        (d) shows the phase $\phi = \arctan p_x/x\gamma^{1/2}$  of the two orbits from (c). 
        The gray shaded area $z>\SI{2.5}{mm}$ indicates the field-free space outside the plasma accelerator. 
        The thick blue line shows the phase difference $\Delta\phi$, which indicates a decreasing decoherence during acceleration.
        The dashed blue line is calculated with equation~(\ref{eq:phase-difference}) using an averaged $\gamma^\prime$.
        The vertical gray dashed line indicates the beginning of the plasma density down-ramp.
        Close to the end of acceleration ($z=\SI{2.0}{\milli\meter}$), 
        (e), (f) and (g) present
        the beam energy bandwidth $\Delta E$,
        the normalized divergence $\gamma \sigma_\Theta$,
        and the normalized beam radius $\sigma_r \gamma^{-1/4}$, respectively. 
	    Error bars indicate one standard deviation in (g).
	    }
    \label{fig:transverse-phase-space-waterfall-plot}
\end{figure}
Figures~\ref{fig:transverse-phase-space-waterfall-plot}(a) and (b) exemplify the evolution of the transverse momentum and beam envelope for the bunch charge of \SI{50}{pC} from injection until extraction to vacuum.
Periodic modulations in the electron beam size are visible, indicating that the bunch is partially coherent during acceleration~\cite{Xu2014} consistent with the estimation of the decoherence length $l_\mathrm{dc}$.
Furthermore, in figure~\ref{fig:transverse-phase-space-waterfall-plot}(c),
the red and green lines represent the orbits of selected particles from the first and last slice of the region II in figure~\ref{fig:injection}(a), respectively.
As shown in figure~\ref{fig:transverse-phase-space-waterfall-plot}(d), the betatron phase $\phi$ and difference $\Delta\phi$ are calculated from $(x,p_x)$. 
After injection, the difference in phase rapidly reduces within \SI{500}{\micro\meter} from an almost fully decoherent bunch ($\Delta\phi \approx \pi$) to a partially coherent bunch ($\Delta\phi \approx \pi/4$).
This is a result of the highly relativistic energy during acceleration which reduces $\omega_\beta$, the rate of increase $\dot{\phi}$, and thus the relative rotation speed of the slices in $(x,p_x)$ with respect to each other.
Partial coherence ($\Delta\phi<\pi$) typically decreases the divergence in the case of a phase space ellipse aligned along the spatial coordinate. 
Therefore, the bunch partly restores coherence during acceleration while starting fully decoherent at injection.
Summarizing the simulation results at the accelerator exit, figure~\ref{fig:transverse-phase-space-waterfall-plot}(e) plots the energy bandwidth where the simulated matched beam loading is obtained at \SI{50}{pC} charge.
Defined by the transverse momentum spread $\sigma_{p}^f$ at the plasma exit,
the normalized divergence $\gamma\sigma_\Theta = \sigma_{p}^f / (m_e c)$ reaches a minimum at this matched beam loading charge, shown 
in figure~\ref{fig:transverse-phase-space-waterfall-plot}(f), consistent with the experiment.
In figure~\ref{fig:transverse-phase-space-waterfall-plot}(g), it can be seen that
the normalized beam size $\sigma_r \gamma^{-1/4}$ does not increase significantly with charge and is about \SI{0.15}{\micro\meter}
which is in agreement with the experimental measurements.

\subsection{Extraction and vacuum propagation}
\begin{table}
    \caption{Space charge effects in the drift space: 
    Beam parameters after \SI{50}{\centi\meter} of free propagation simulated using GPT.}
    \begin{tabular*}{0.75\textwidth}{r|ccccc}
        charge & norm. & geom. & beam & average & energy \\
        & divergence  & divergence & size & energy & spread\\
         {[pC]}  & {[rad]} & {[mrad]} & {[mm]} & {[\si{MeV}]} & {[\si{MeV}]} \\
        \hline
        100 & 3.29 & 4.4  & 2.2  & 380 & 20    \\
        300 & 3.34 & 4.4  & 2.2  & 389 & 22    \\
        500 & 3.49 & 4.4  & 2.2  & 397 & 24.5 \\
    \end{tabular*}
\label{tab:beam-in-drift-space}
\end{table}
In our experiment, 
accelerated electrons are detected after propagation in a vacuum much longer than the plasma channel
and space charge effects could affect the measured beam parameters.
For an efficient simulation of the beam dynamics in such a drift space,
 particle tracing simulations were performed with the General Particle Tracer (GPT) code~\cite{GPT}, 
taking the space charge effect via the SpaceCharge3D-model into account.
From the previously performed PIC simulations as shown in figure~\ref{fig:transverse-phase-space-waterfall-plot}, 
the macro-particles at the plasma exit were extracted, reduced to a set of $5000$ and fed into GPT~\cite{Ksenia2021}.
The charge of this particle distribution was up-scaled to \SIlist{100;300;500}{\pico\coulomb}
in order to mimic experimentally observed maximal charges and to accurately study the space charge effect.
Reflecting experimental conditions, these macro-particles were tracked from the plasma exit throughout a \SI{50}{\centi\meter} long drift space.
Table~\ref{tab:beam-in-drift-space} lists the transverse beam parameters of a typical electron bunch from these simulations. 
Energy and energy spread increase by \SI{5}{\percent} and \SI{20}{\percent} for higher bunch charges, respectively.
The beam size and geometrical divergence after propagation seem to be charge-independent
and the normalized divergence monotonically increases by about \SI{10}{\percent} when the charge increases by a factor of five.
For a high bunch charge, electrons are accelerated by the Coulomb repulsion which causes electrons to gain transverse and longitudinal momentum~\cite{Gruner2009}.
The monotonic increase is in contrast to the nonlinear charge-dependent betatron decoherence described before.
Thus, space charge effects during propagation in the drift space cannot explain the minima in the normalized divergence.

\section{Summary}
In summary, we experimentally demonstrated that matched beam-loading in a laser-wakefield accelerator, 
identified via its characteristic charge-dependent minimum in beam energy spread, 
additionally yields a minimum in normalized beam divergence.
The experiment relied on control over the injected charge in the self-truncated ionization injection regime and on monitoring the beam diameter inside the plasma via betatron x-ray spectroscopy.
Combined with simplified analytical modeling of the betatron phase advance per given acceleration length, 
the experiment suggests that the observation of reduced divergence is a result of the suppression of betatron phase decoherence. 
Detailed studies of the complete six-dimensional phase space dynamics of injected electrons using high-resolution particle-in-cell simulations revealed that despite the injection extending over the full initial betatron period, the thus decoherent bunch is partially restoring its betatron coherence during acceleration in the beam-loaded field.

This finding substantiates the importance of operating plasma accelerators in the matched beam-loading regime for the simultaneous optimization of beam charge and transverse emittance. 
High beam quality and in particular low divergence is essential for the realization of the next generation of high-brightness light sources and facilitates beam transport to subsequent stages or insertion devices like undulators.
In combination with the high-peak currents of the beam-loaded regime, this demonstration of optimized electron beam
quality opens a pathway to the realization of high-brightness radiation sources.

The authors gratefully acknowledge the team from DRACO as well as 
T. Cowan, M. Downer, J. Grenzer, A. Laso Garc\'ia, 
K. Steiniger, A. Wagner and R. Zgadzaj for their support. 
This work was partly supported by EuCARD-2, funding under Grand Agreement No. 312453, 
by HZDR under programme Matter and Technology,
and by the Center of Advanced Systems Understanding (CASUS) which is financed by Germany’s Federal Ministry of Education and Research (BMBF) and by the Saxon Ministry for Science, Culture and Tourism (SMWK) with tax funds on the basis of the budget approved by the Saxon State Parliament.
The authors gratefully acknowledge the GWK support for funding this project by providing computing time through the Center for Information Services and HPC (ZIH) at TU Dresden on the HRSK-II.
\FloatBarrier
\newpage
\appendix
\section*{Appendix}
\setcounter{section}{1}
\label{app:experimental-figures}

\subsection*{Supplementary experimental plots}
\FloatBarrier
\begin{figure}[h]
 \centering
    \includegraphics[width=0.45\textwidth]{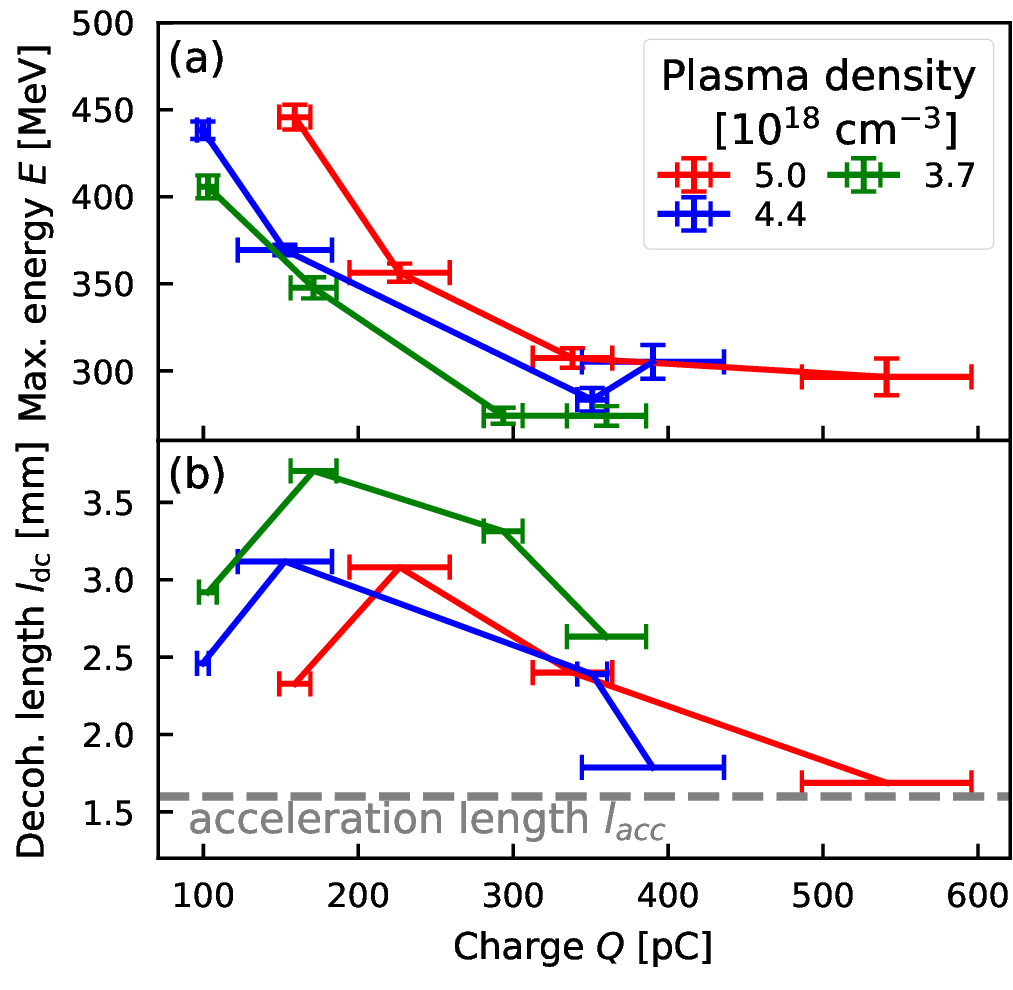}
    \caption{(a) shows the maximum attained energy $E$
    and (b) presents the betatron decoherence length 
    $l_\mathrm{dc} =  2.35 \lambda_\beta \langle E \rangle /\Delta E$.}
    \label{fig:duoplot-over-charge-supplement}
\end{figure}
Figure~\ref{fig:duoplot-over-charge-supplement}(a) presents the maximum electron energy of the accelerated electron bunch.
Each data points has up to $15$ consecutive shots and  error bars represent the standard error of the mean.
The energy decreases with charge due to the flattened accelerating field.
Figure~\ref{fig:duoplot-over-charge-supplement}(b) shows the betatron decoherence length $l_\mathrm{dc}$ required for full decoherence of the beam.
The acceleration length $l_\textrm{acc}$ obtained from simulations is shorter than $l_\mathrm{dc}$.
The maximum in $l_\mathrm{dc}$ is consistent with the minimum phase difference shown in figure~\ref{fig:experimental-results}(d).
The total plasma length including \SI{0.5}{\milli\meter} up- and down-ramps is \SI{2.6}{\milli\meter}
and thus the maximum $l_\mathrm{dc}$ is not reached in the experiment.

\subsection*{PIC simulations}
\label{app:PIC}
The PIC simulations
utilized a laser pulse that was modeled using a Gauss-Laguerre envelope based on experimental measurements in transverse space and a Gaussian envelope in time.
The laser wavelength was $\lambda_0=\SI{800}{\nano\meter}$ 
and a pulse length was \SI{30}{\femto\second} (full-width at half-maximum).
The simulation box consisted of \num{768 x 4608 x 768} cells 
and had a transverse resolution of $0.33\lambda_0$,
a longitudinal resolution of $0.0283\lambda_0$ and a temporal resolution of $0.0273 c/\lambda_0$.
Such spatio-temporal resolution is required 
in order to numerically investigate the dynamics of the wakefield and plasma electrons.
The electromagnetic fields and macro-particles propagation were compiled with
the field solver by Lehe~\cite{Lehe2016} and the particle pusher by 
Boris~\cite{Boris1970}, 
respectively.
The current was calculated with the Esirkepov deposition scheme~\cite{Esirkepov2001} and a triangular shaped density cloud interpolation~\cite{Hockney1988}.
The simulations utilized a combined BSI~\cite{Mulser2010} and ADK~\cite{Delone1998} method for ionization implementation.
All simulations were performed with particle identifiers that allowed particle tracking of the nitrogen K-shell electrons.

The slightly lower charge obtained in simulations than in the experiment can be a result of idealizations in the laser and gas modeling. 
Supported by the higher electron energy in simulations, a possible explanation is that electrons are injected at higher accelerating fields.
For a given plasma density, 
Tzoufras et al.~\cite{Tzoufras2008} found that the optimal loading charge is inverse proportional to the accelerating field.
Thereby, the idealized laser parameters causes a stronger acceleration and thus a reduced beam loading charge.

\begin{figure}[h]
    \centering
    \includegraphics[width=\textwidth]{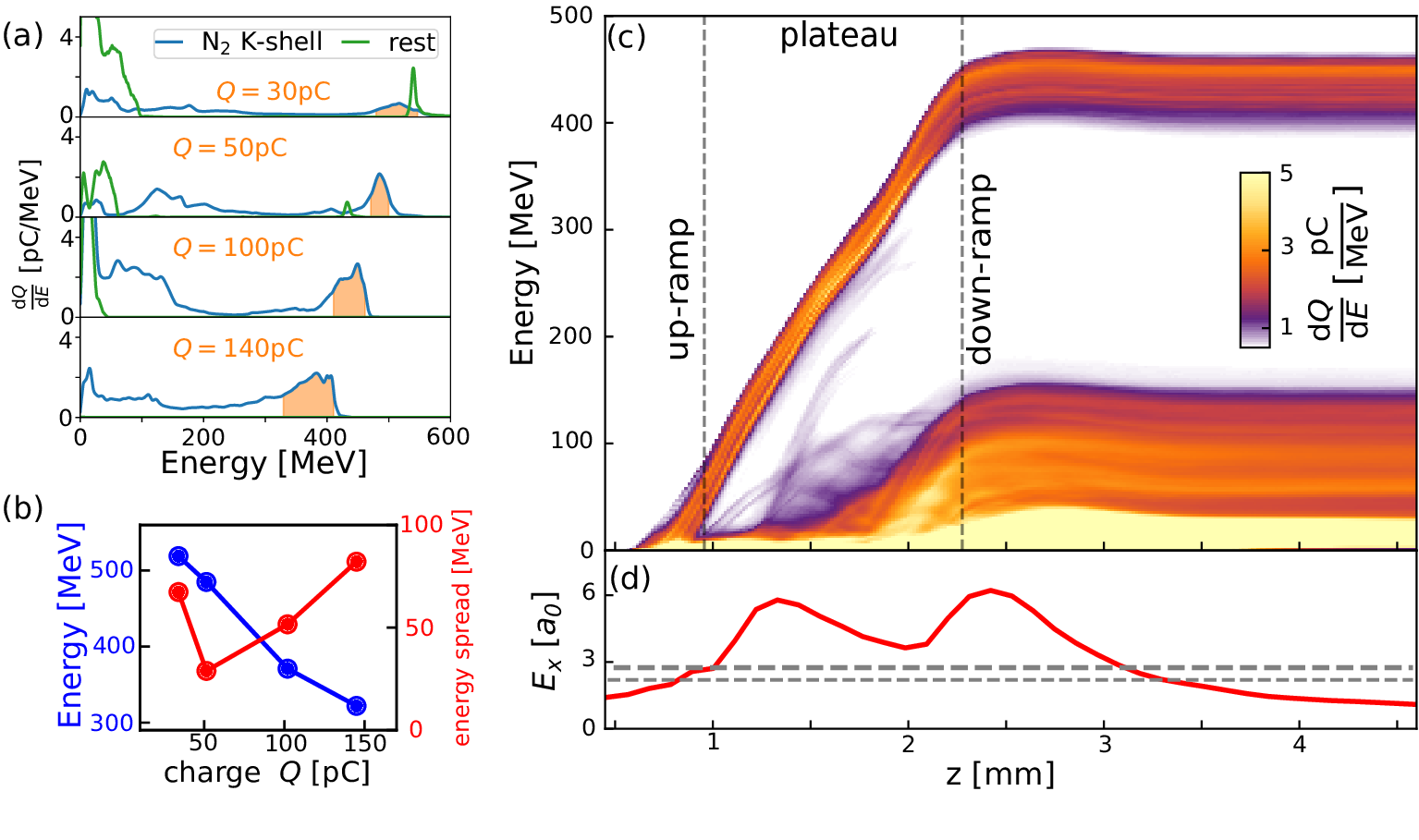}
    \caption{{PIC simulations with different dopings:}
	(a) shows the electron energy histograms of four simulations with \SIlist{0.25;0.5;1;3}{\percent} nitrogen doping.
	(b) illustrates the charge dependency of energy and energy spread of the simulations shown in (a).
	(c) presents the electron energy evolution during the acceleration for the simulation with \SI{1}{\percent} nitrogen doping.
	(d) shows the on-axis laser strength $a_0$,
	with the gray dashed lines indicating the required field strength for ionization of the first and second K-shell electron of nitrogen.
	The plasma density is \SI{4.4}{\times10^{18}\centi\meter^{-3}}.}
    \label{pix:pic-spectra}
\end{figure}
Figure~\ref{pix:pic-spectra} presents the simulation data of the doping scan. 
The electron spectra at the end of the simulation is shown at figure~\ref{pix:pic-spectra}(a).
From the histogram of the K-shell electrons, 
the maximum electron energy and energy spread within full-width at half-maximum (FWHM) 
is plotted for the charge (FWHM) in figure~\ref{pix:pic-spectra}(b).
Energy and energy spread clearly shows beam loading as discussed in the manuscript.
The energy decrease is consistent with the experimental data shown in figure~\ref{fig:duoplot-over-charge-supplement}(a).
Figure~\ref{pix:pic-spectra}(c) presents the typical evolution of electron energy and normalized laser intensity $a_0$.
All simulations indicate an injection at $z=\SI{1}{\milli\meter}$ over a distance of \SI{300}{\micro\meter}.

\begin{figure}[b!]
    \centering
    \includegraphics[width=0.8\textwidth]{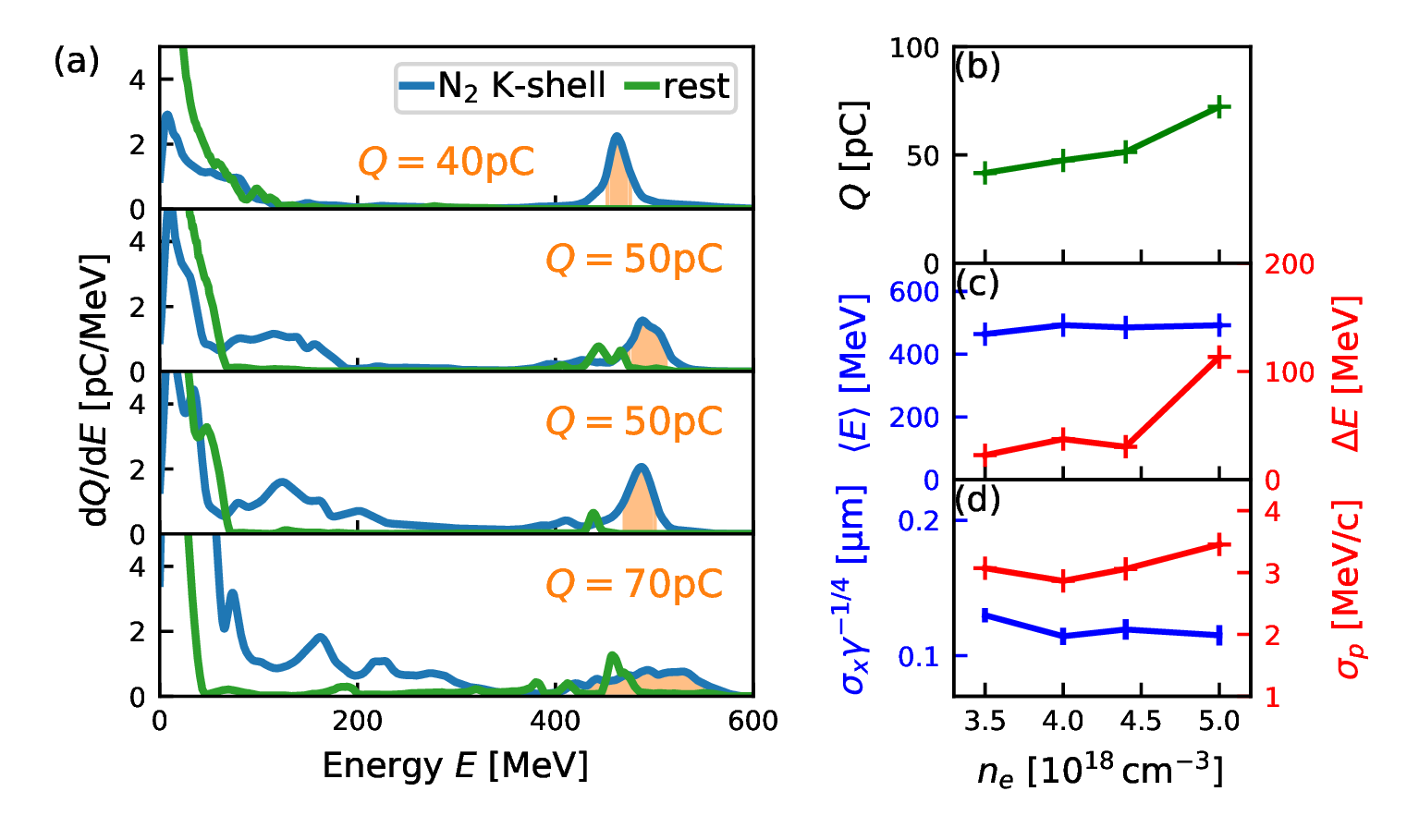}
    \caption{{PIC simulations with different plasma densities:}
	(a) shows the electron energy histograms of four simulations with plasma densities $n_e$ of \SIlist{3.5;4.0;4.4;5.0}{\times 10^{18}\centi\meter^{-3}}.
	(b) and (c) illustrate the plasma density dependency of energy, energy spread and charge of the simulations shown in (a).
	(d) presents the normalized beam size $\sigma_x \gamma^{-1/4}$ and transverse momentum spread $\sigma_p$ over different densities.
	}
    \label{pix:pic-spectra-density-scan}
\end{figure}
Figure~\ref{pix:pic-spectra-density-scan} shows the simulation data of a density scan at a nitrogen doping of \SI{0.5}{\percent}.
The charge within FWHM weakly depends on the density.
As theoretically expected, the energy increases with plasma density and the energy spread does not depend on the density. 
The normalized beam size is independent of the density, as observed experimentally.
The transverse momentum spread increases with plasma density, which was also the case in the experiment.

\FloatBarrier
\newpage
\bibliographystyle{unsrt}
\bibliography{library}

\end{document}